\documentstyle{ichep}
\def\.{\!\cdot\!}
\def\tr{{\rm tr}}
\def\p{\partial}
\def\bk#1{\langle#1\rangle}

\begin{document}

\title{String rearrangement of gauge theories}

\author{C.S. Lam \dag }

\affil{Department of Physics, McGill University\\
3600 University St., Montreal, P.Q., Canada H3A 2T8}

%\collab{333}

\abstract{Feynman diagram expressions in ordinary field theories can
be written in a string-like manner. The methods and the advantages for
doing so are briefly discussed.}

\twocolumn[\maketitle]

\fnm{1}{E-mail: lam@physics.mcgill.ca}

\section{Introduction}
There are three reasons why one wants to arrange ordinary Feynman diagrams
in a string-like manner: it simplifies calculations and it gives new insights
into gauge and gravitational theories. Moreover, this is done (graphically) all
the times so we might as well find out exactly what that means. For example,
The tree diagrams for the process
$\pi^+K^0\to \pi^0K^+$
are given by Figs.~1(a) and (b), but  one often
 shows only the quark diagram Fig.~1(c), which can  be considered also as
a string diagram with hadronic strings strung between  $q\bar q$ pairs.
String diagrams are also
used for pure QCD in the large-$N_c$ limit as shown in
Fig.~2.

\begin{figure}
\vspace*{11.5cm}
\caption{Lowest-order diagrams for $\pi^+K^0\to \pi^0K^+$.}
\end{figure}

\begin{figure}
\vspace*{9.5cm}
\caption{A pure-gluon QCD process.}
\end{figure}

Is it possible then to make the string-like diagrams {\it quantitative}
 by writing down for them a set of `Feynman rules'? When one tries to
do that a number of problems is encoutered.
For ordinary Feynman diagrams,
($i$) loop momenta $k_a$ have to be introduced and integrated over;
($ii$) interactions occur at the vertices though particles
propagate freely between them. In particular, loop momenta $k_a$
injected at the vertices can
change the direction of
the combined momentum flow. Similarly, flavour, colour, and spin are altered
at the vertices; ($iii$) gauge invariance
determines the vertex factor
for gauge interactions, for example to be
$\epsilon(p)\cdot(q'+q'')$ in scalar QED, where $p$ is
the momentum of the external photon  and $q',q''$ are the charged
particle momenta; ($iv$)
a sum of many Feynman diagrams is needed to
describle a process in a given order. In contrast, for string diagrams,
($i'$) if we consider only hadronic {\it ground} states so as to freeze
string excitations,  the propagation of a string is described
only by the `proper time' parameter $\tau$, so that scattering amplitudes
are expressed as integrations over
the proper-time parameters $\tau_m$;
($ii'$) the geometrical shape of a string diagram is arbitrary
(reparametrization and conformal invariance), so
vertices cannot be present and
strings propagate freely throughout the diagram. In particular, external
momenta can merge and divide inside a diagram but there are no
additional loop momenta
present to alter their directions. Flavour and colour flow smoothly along the
quark lines
as given respectively by Fig.~1(c)
and Fig.~2(b), and similarly for spin as we shall see later;
($iii'$) conformal invariance
determines external gauge-interaction vertices, for example to be
$\epsilon(p)\cdot[\partial_\tau x(\tau)]\exp[ip\cdot x(\tau)]$ for scalar
QED
where $x^\mu(\tau)$ is the spacetime (operator) coordiantes of the string
at the proper time $\tau$. The vertex factor for this operator is then
$\epsilon(p)\cdot[\partial_\tau(-i\partial/\partial p)]$,
quite different from the corresponding field-theoretic vertex in ($iii$);
($iv'$) (Veneziano) duality [1] is valid. The sum of many Feynman diagrams
is replaced by one or few string diagrams, as in Fig.~1.

In the rest of this note, we will show how to rearrange
($i$)--($iv$) to resemble ($i'$)--($iv'$). The calculational and
conceptual advantages for doing so will also be briefly discussed.

\section{Momentum, flavour, colour, and spin flows}
To convert ($i$) to ($i'$), it is sufficient to introduce a Schwinger
parameter $\alpha_r$ for the denominator of every propagator:
\begin{equation}
(-q_r^2+m_r^2-i\epsilon)^{-1}=i\int_0^\infty d\alpha_r\exp[-i\alpha_r
(m_r^2-q_r^2)]\ .
\end{equation}
Substituting this into the general expression for a T-matrix
amplitude
\begin{equation}
T(p)=\int\left(\prod_ad^4k_a\right)S_0(q,p)\prod_r(-q_r^2+m_r^2-i\epsilon)^{-1}
\ ,
\end{equation}
the loop momentum integrations can be explicitly carried out to yield [2]
\begin{eqnarray}
T(p)\sim&&\int_0^\infty\left(\prod_rd\alpha_r\right)\Delta^{-2}(\alpha)
S(q,p)\nonumber\\
&&\prod_r\exp[-i\alpha_r(m_r^2-q_r^2)]\ .\\
\nonumber
\end{eqnarray}
This is the string-like form where momentum flow
is described by $q_r$, a quantity best thought of as
the current flowing through
the $r$th line of an electric circuit
given by the Feynman diagram, in which
the branch resistances are $\alpha_s$ and the external currents
are $p_i$. Explicit rules are available
to calculate these currents and other quantities in $T(p)$
directly from the Feynman diagram. If a proper time $\tau$ is assigned
to each vertex, and if line $r=(ij)$ connects vertices $j$ to $i$, then
$\alpha_r$ can also be interpreted as the proper-time difference
$|\tau_i-\tau_j|$.

The factor $S_0(q,p)$ in (2) contains all the vertex factors and
numerators of propagators, so it encodes
the flows of flavour, colour, and spin and is given by a sum of products
of these factors. The quantity $S(q,p)$ in (3) is equal to
$S_0+S_1+S_2+\cdots$, and can be obtained from $S_0$ through momentum
contraction [2]. The factors for flavour, colour, and spin flows can all
be read off directly from the appropriate quark, or string-like, diagrams,
and it is important to note that the quark diagrams are {\it different for
different flows}. The general rules and their derivations are given elsewhere
[3], but specific examples can be seen from Figs.~3 and 4 respectively for
colour and spin flows. In each case diagram (a) is the Feynman diagram
and diagram (b) is the equivalent string-like diagram from which
the factors for $S_0$ can be read off to be

\begin{figure}
\vspace*{9cm}
\caption{An example to illustrate colour flow. See eq.~(4).}
\end{figure}

\begin{figure}
\vspace*{10cm}
\caption{An example to illustrate spin flow. See eq.~(5).}
\end{figure}

\begin{eqnarray}
S_0^{(m)colour}=&&T^aT^bT^cT^d \tr(T^eT^fT^gT^h)\ ,\\
S_0^{spin}=&&[p_2q_1]\bk{q_1p_3}\.[p_4p_5]\.
\bk{k_5q_9}[q_9q_3]\bk{q_3p_1}\. \nonumber\\
&&\bk{p_6q_6}[q_6q_5]\bk{q_5q_8}[q_8k_6]\ .\\
\nonumber
\end{eqnarray}
$T^a$ in (5) are the $(S)U(N)$ generators in the fundamental representation.
The superscript $(m)$ indicates that there are many colour
factors for Fig.~3(a), and that in (4) is the one appropriate to
Fig.~3(b). Other colour factors can be obtained from
other string-like diagrams, which in the case of $U(N)$ can be obtained from
3(b) by crossing
the quark lines at the vertices. For $SU(N)$ there are other quark diagrams
in which some of the internal gluon lines are omitted.
The spin-flow (or more correctly helicity-flow) factor $S_0^{spin}$ in (5) is
obtained by assuming
the fermion masses to be zero. In that case the string-like diagram 4(b)
is unique once 4(a) is given, and the direction the fermion lines turn depends
on the helicities of the external particles, indicated in the diagram by a
$+$ or a $-$ sign. More diagrams will be necessary if the internal
particles are massive. The square and angular brackets are the overlap of
the massless Dirac wave functions with definite helicities:
\begin{equation}
[p_ip_j]=\bar u_+(p_i)u_-(p_j),\quad \bk{p_ip_j}=\bar u_-(p_i)u_+(p_j)\ ,
\end{equation}
When internal momenta $q_r$ appear in these brackets, it is understood that
they should first be expanded in terms of the external massless momenta
$p_i$ with only diagonal terms kept. Note that Dirac matrices are completely
absent so the usual four-channel problem is reduced to a one-channel
expression. This is possible because of helicity and chirality conservations
for massless fermions. Note that it would have been impossible to write
(5) before the loop momenta were eliminated in going from
(2) to (3).

The final factor $S_0$ is obtained from
$S_0^{(m)colour}S_0^{spin}\cdots$ by summing over all $m$. The main
advantage in this string-like arrangement is to be able to use the spinor
helicity technique [4], developed originally for {\it tree} diagrams, now  for
arbitrary processes with any number of loops. It also allows
the gauge-invariant colour subamplitudes to be easily separated.

\section{External gauge vertices}
The scalar QED vertex ($iii$) can be replaced by the string-like vertex
($iii'$) by using differential circuit identities [3]. The presence of
$\p/\p\tau$ in the latter expression allows integration-by-parts
to be used, to redistribute the gauge-dependent terms among
different diagrams in order to minimize their appearance and thereby increase
computability. It also makes the Ward-Takahashi identity realized
in a different way, and allows the possibility of formulating gauge
invariance in another way.

\section{Duality}
Veneziano duality ($iv'$) can be simulated by formally summing up a number
of Feynman diagrams into a single integral expression, at least for
QED-like theories [5]. Take for
example the sum of the $6!4!3!$ QED diagrams,
obtained from Fig.~5 by permuting the photon vertices along each of the
three charged lines.

\begin{figure}
\vspace*{5 cm}
\caption{A QED diagram.}
\end{figure}

In the proper-time representation (3),
a proper time [$\tau'_i(1\le i\le 6),\ \tau''_j(1\le j\le 4)$, \
$\tau'''_k(1\le k\le 3)$] is assigned to each vertex, and each Feynman
diagram is given by a fixed ordering of the $\tau'$s, $\tau''$s,
and $\tau'''$s, so for each diagram the proper-time integration region in (3)
is the product of three {\it hyper-triangular regions}. The `dual {\it sum}' of
these $6!4!3!$
diagrams is obtained by summing all these permutations, and is thus
given by a {\it single} integral over the product of three
{\it hypercube} regions, one each for $\tau',\tau''$,  and $\tau'''$.
Moreover, it is gauge invariant. Unfortunately, it is often impossible
to carry out this single integral analytically, but it can be used as
a starting point for gauge-invariant approximations. For example, the
result for the soft-photon eikonal approximation can be obtained very easily
in this way.

It is also worthwhile pointing out an interesting parallel. Each Feynman
diagram with $n$ vertices is a sum of $n!$ time-ordered (old-fashioned)
diagrams. An individual time-ordered diagram is not Lorentz invariant,
but the Feynman diagram is. In our case, the dual sum puts together
Feynman diagrams that have different {\it proper-time} orderings, each
of which is not gauge invariant, but the dual sum is.

\section{References}
\def\i#1{\item{[#1]}}
\def\npb#1{{\it Nucl.~Phys. }{\bf B#1}}
\def\plb#1{{\it Phys.~Lett. }{\bf #1B}}
\def\prl#1{{\it Phys.~Rev.~Lett. }{\bf B#1}}
\def\prd#1{{\it Phys.~Rev. }{\bf D#1}}
\def\pr#1{{\it Phys.~Rep. }{\bf #1}}
\def\ibid#1{{\it ibid.} {\bf #1}}

%\Bibliography

\end{document}